# Validating and Updating GRASP: A New Evidence-Based Framework for Grading and Assessment of Clinical Predictive Tools

Mohamed Khalifa [1], Farah Magrabi [1] and Blanca Gallego [2]

[1] Australian Institute of Health Innovation, Faculty of Medicine and Health Sciences, Macquarie University, Sydney, Australia

[2] Centre for Big Data Research in Health, Faculty of Medicine, University of New South Wales, Sydney, Australia

**Authors:**

Mohamed Khalifa: mohamed.khalifa@mq.edu.au

Farah Magrabi: farah.magrabi@mq.edu.au

Blanca Gallego: b.gallego@unsw.edu.au

**Abstract**

**Background:** When selecting predictive tools, for implementation in their clinical practice or for recommendation in clinical guidelines, clinicians are challenged with an overwhelming and ever-growing number of tools. Many of these have never been implemented or evaluated for comparative effectiveness. To overcome this challenge, the authors developed an evidence-based framework for grading and assessment of predictive tools (GRASP), based on the critical appraisal of their published evidence. The objective of this study is to validate, update GRASP, and evaluate its reliability.

**Methods:** The study is composed of two parts. The first includes validating and updating the GRASP framework and the second includes evaluating the framework reliability. For the first part, an online survey was developed to collect the responses of a wide international group of experts; identified as healthcare researchers who have published studies on developing, implementing or evaluating predictive tools and clinical decision support systems. For the second part, the interrater reliability of the framework, to assign grades to eight predictive tools by two independent users, will be evaluated.

**Results:** Out of 882 invited experts, 81 valid responses were received. On a five-points Likert scale, experts overall strongly agreed to GRASP evaluation criteria of predictive tools (4.35/5). Experts strongly agreed to six criteria; predictive performance (4.87/5) and predictive performance levels (4.44/5), usability (4.68/5) and potential effect (4.61/5), post-implementation impact (4.78/5) and evidence direction (4.26/5). Experts somewhat agreed to one criterion; post-implementation impact levels (4.16/5). Experts were neutral about one criterion; usability is higher than potential effect (2.97/5). Two thirds of the experts provided recommendations to six open-ended questions regarding adding, removing or changing evaluation criteria. Over half of the experts suggested that the potential effect, as an evaluation criterion, should be higher than the usability. Experts highlighted the importance of reporting the quality of studies and the strength of evidence supporting the grades assigned to predictive tools. The GRASP concept and its detailed report were updated based on experts' feedback. Following validation, the interrater reliability of the GRASP framework, to produce accurate and consistent results by two independent users, was tested and the framework found to be reliable. Answering open-ended questions, the two independent users reported the GRASP framework was logical, useful, and easy to use.

**Discussion and Conclusion:** The GRASP framework grades predictive tools based on the critical appraisal of the published evidence across three dimensions: 1) Phase of evaluation; 2) Level of evidence; and 3) Direction of evidence. The final grade of a tool is based on the highest phase of evaluation, supported by the highest level of positive evidence, or mixed evidence that supports positive conclusion. GRASP is not meant to be prescriptive; it provides clinicians with a high-level, evidence-based, and comprehensive, yet simple and feasible, approach to evaluate and compare clinical predictive tools, considering their predictive performance before implementation, potential effect and usability during planning for implementation, and post-implementation impact on healthcare and clinical outcomes.

## 1. Background

Clinical decision support (CDS) systems have been proved to enhance evidence-based clinical practice and support healthcare cost-effectiveness [1-6]. According to the definition developed by Edward Shortliffe, there are three levels of CDS functions, these include; 1) managing health information through providing tools for search and retrieval, 2) focusing users' attention through flagging abnormal values or possible drug-to-drug interactions, and 3) providing patient specific recommendations based on the clinical scenario, which usually follow rules and algorithms, cost-benefit analysis or clinical pathways [7, 8]. Clinical predictive tools, here referred to simply as predictive tools, belong to the third level of CDS and include various applications; ranging from the simplest manual clinical prediction rules to the most sophisticated machine learning algorithms [9, 10]. These research-based applications provide diagnostic, prognostic, or therapeutic decision support. They quantify the contributions of relevant patient characteristics to derive the likelihood of diseases, predict their courses and possible outcomes, or support the decision making on their management [11, 12].

When selecting predictive tools, for implementation in their clinical practice or for recommendation in clinical practice guidelines, clinicians involved in the decision making are challenged with an overwhelming and ever-growing number of tools. Many of these tools have never been implemented or assessed for comparative performance or impact [13-15]. Currently, clinicians rely on their previous experience, subjective evaluation or recent exposure to predictive tools in making selection decisions. Objective methods and evidence based approached are rarely used in such decisions [16, 17]. Some clinicians, especially those developing clinical guidelines, search the literature for best available published evidence. Commonly they look for research studies that describe the development, implementation or evaluation of predictive tools. More specifically, some clinicians look for systematic reviews on predictive tools, comparing their predictive performance or development methods. However, there are no available approaches to objectively summarise or interpret such evidence [18, 19].

### 1.1. The GRASP Framework

To overcome this major challenge, the authors have developed a new evidence-based framework for grading and assessment of predictive tools (The GRASP Framework) [20]. The aim of this framework is to provide clinicians with standardised objective information on predictive tools to support their search for and selection of effective tools for their tasks. Based on the critical appraisal of the published evidence on predictive tools, the GRASP framework uses three dimensions to grade predictive tools: 1) Phase of Evaluation, 2) Level of Evidence and 3) Direction of Evidence.

**Phase of Evaluation:** Assigns A, B, or C based on the highest phase of evaluation. If a tool's predictive performance, as reported in the literature, has been tested for validity, it is assigned phase C. If a tool's usability and/or potential effect have been tested, it is assigned phase B. Finally, if a tool has been implemented in clinical practice, and there is published evidence evaluating its impact, it is assigned phase A.

**Level of Evidence:** A numerical score, within each phase, is assigned based on the level of evidence associated with each tool. A tool is assigned grade C1 if it has been tested for external validity multiple times; C2 if it has been tested for external validity only once; and C3 if it has been tested only for internal validity. C0 means that the tool did not show sufficient internal validity to be used in clinical practice. Similarly, B1 is assigned to a predictive tool that has been evaluated during implementation for its usability; while if it has been studied for its potential effect on clinical effectiveness, patient safety or healthcare efficiency, it is assigned B2. Finally, if a predictive tool had been implemented then evaluated after implementation for its impact, on clinical effectiveness, patient safety or healthcare efficiency, it is assigned score A1 if there is at least one experimental study of good quality evaluating its impact, A2 if there are observational studies evaluating its impact and A3 if the impact has been evaluated through subjective studies, such as expert panel reports.

**Direction of Evidence:** For each phase and level of evidence, a direction of evidence is assigned based on the collective conclusions reported in the studies. The evidence is considered positive if all studies about a predictive tool reported positive conclusions and negative if all studies reported negative or equivocal conclusions. The evidence is considered mixed if some studies reported positive and some reported either negative or equivocal conclusions. To decide an overall direction of evidence, a protocol is used to sort the mixed evidence into supporting an overall positive conclusion or supporting an overall negative conclusion. The protocol is based on two main criteria; 1) The degree of matching between the evaluation study conditions and the original predictive tool specifications, and 2) The quality of the evaluation study. Studies evaluating predictive tools in closely matching conditions to the tool specifications and providing high quality evidence are considered first for their conclusions in deciding the overall direction of evidence. The mixed evidence protocol is detailed and illustrated in Figure 6 in the Appendix.

The final grade assigned to a tool is based on the highest phase of evaluation, supported by the highest level of positive evidence, or mixed evidence that supports a positive conclusion. The GRASP framework concept is shown in Figure 1 and the GRASP framework detailed report is presented in Table 3 in the Appendix.

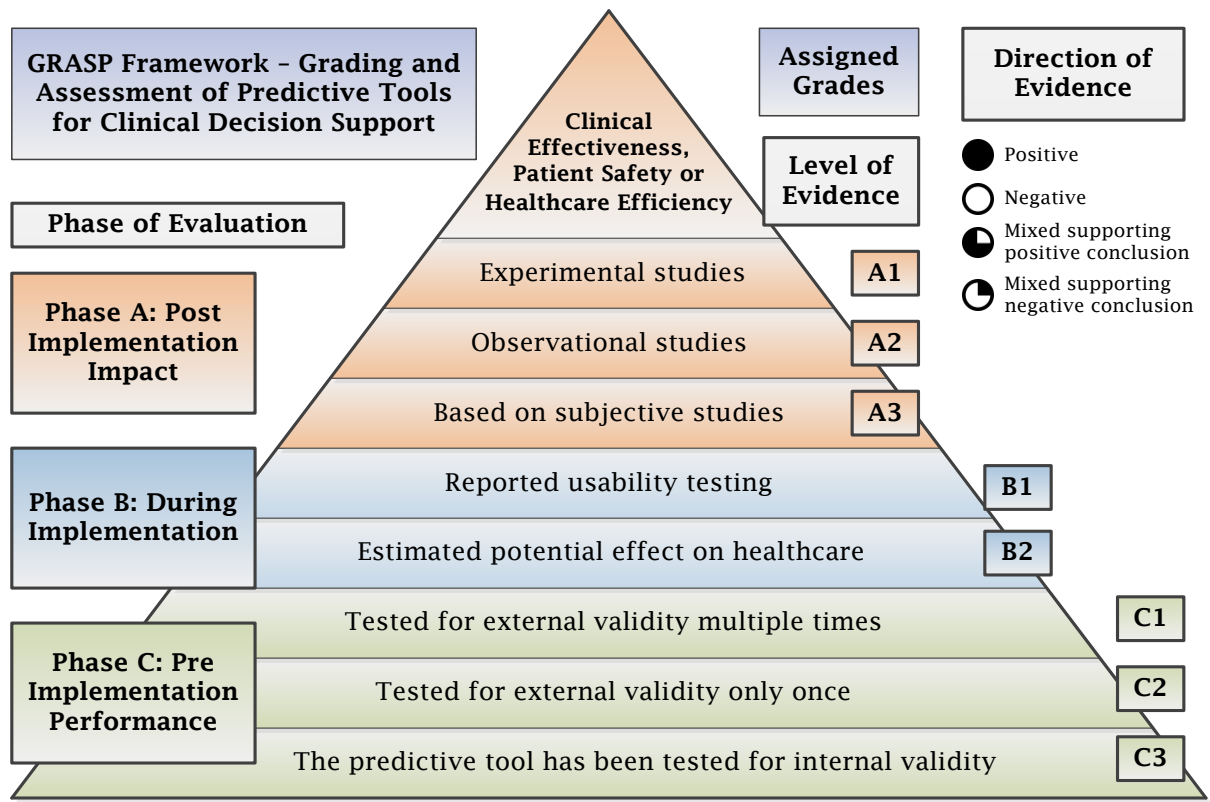

Figure 1: The GRASP Framework Concept [20]

*1.2. Study Objectives*

Validating new clinical instruments, healthcare models and evaluation frameworks through the feedback of experts is a well-established approach, especially in the area of CDS [12, 21, 22]. Using a mixed approach of qualitative and quantitative methods in research proved to be useful in healthcare, because of the complexity of the studied topics [23]. Using open-ended questions in quantitative surveys adds significant value and depth to both the results and conclusions of studies conducted [24]. The aim of this study is to validate and update the GRASP framework and to evaluate its reliability. The primary objective is to validate and update the criteria used by the GRASP framework, for grading and assessment of predictive tools, through the feedback of a wide international group of healthcare experts in the areas of developing, implementing and evaluating clinical decision support systems and predictive tools. The secondary objective is to evaluate the GRASP framework reliability to ensure that the outcomes produced by independent users, when grading predictive tools using the GRASP framework, are accurate, consistent and reliable.

## 2. Methods

### *2.1. The Study Design*

The study is composed of two parts. The first part includes validating and updating the GRASP framework and the second part includes evaluating the framework reliability. For the first part, a survey was designed to solicit the feedback of experts on the criteria used by the GRASP framework for grading and assessment of predictive tools. The main outcome of this part of the study is to measure the validity and update the design and content of the GRASP framework. The analysis includes evaluating the degree of agreement of experts on how essential the different criteria used to grade predictive tools are, including the three dimensions; phases of evaluation (before, during and after implementation), levels of evidence and directions of evidence within each phase. In addition, experts' feedback on adding, removing or updating any of the criteria, used to grade predictive tools, and their further suggestions and recommendations will also be analysed and considered.

Based on similar studies; validating and updating systems through surveying expert users, it was estimated that the required sample size for this study is around fifty experts [25-28]. Experts were identified as researchers who have published at least one paper on developing, implementing or evaluating predictive tools and clinical decision support systems. To search for such publications, the concepts of Clinical Decision Support, Clinical Prediction, Developing, Validating, Implementing, Evaluating, Comparing, Reviewing, Tools, Rules, Models, Algorithms, Systems, and Pathways were used. The search engines used were MEDLINE, EMBASE, CINAHL and Google Scholar. For the purpose of emails currency, the search was restricted to the last three years.

The study has been approved by the Human Research Ethics Committee, Faculty of Medicine and Health Sciences, Macquarie University, Sydney, Australia, on the 4[th] of October 2018. The authors expected the distribution of the survey to take two weeks, and the collection of the feedback to take another four weeks. The authors expected the response rate to be around 10%. Before the deployment of the survey a pilot testing was conducted through asking ten experts to take the survey. The feedback of the pilot testing was used to improve the survey design and content; some questions were rephrased, some were rearranged, and some were supported by definitions and clarifications. Experts who participated in the pilot testing were excluded from the participation in the final survey. An invitation email, introducing the study objectives, the survey completion time, which was estimated at 20 minutes, and a participation consent was submitted to the identified experts with the link to the online survey. A reminder email, in two weeks, was sent to the experts who have not responded or completed the survey.

### *2.2. The Study Survey*

The online survey was developed using Qualtrics experience management platform [29]. The online survey, as illustrated through the screenshots in the Appendix, included eight five-points Likert scale closed-ended agreement questions and six open-ended suggestions and recommendations questions distributed over seven sections. The introduction informed the participants about the aim of developing the GRASP framework, its design, and the task they are requested to complete. In addition to informing them that they can request feedback and acknowledgement as well as providing them with contacts to ask for further information or to make complaints. The second section asked experts about their level of agreement with the evaluation of the published evidence on the tools' predictive performance before implementation, such as internal and external validation. The third section asked experts about their level of agreement with the evaluation of the published evidence on the tools' usability and/or potential effect on healthcare during implementation. The fourth section asked experts about their level of agreement with the evaluation of the published evidence on the tools' post-implementation impact on clinical effectiveness, patient safety or healthcare efficiency. The fifth section asked experts about their level of agreement with the evaluation of the direction of the published evidence on the tools, being positive, negative or mixed. Experts were also requested to provide their free-text feedback on adding, removing or changing any of the criteria used for the assessment of phases of evaluation, levels of evidence, or directions of evidence. The sixth section asked experts to provide their free-text feedback on the best methods suggested to define and capture successful tools' predictive performance, when different clinical prediction tasks have different predictive performance requirements. Experts were also asked about managing conflicting evidence of studies when there is variability in the quality and/or sub-populations of the published evidence.

### *2.3. Reliability Testing*

The second part of this study; evaluating the framework reliability, followed the completion of the first part and used the validated and updated version of the GRASP framework. Two independent and experienced researchers were trained, for four hours each by the authors, on using the framework to grade predictive tools. The researchers were then asked to grade eight different predictive tools independently, using the GRASP framework, the full text studies describing the development of the tools, and the comprehensive list of all the published evidence on each tool along with the full text of each study. The objective of this part of the study was to measure the reliability of using the framework, by independent users, to grade predictive tools. Since the tested function of the GRASP framework here is grading tools, the interrater reliability was the best measure to evaluate its reliability. The interrater reliability, also called interrater agreement, is the degree of agreement or the score of how much homogeneity, or consensus, there is in the ratings given by independent judges [30]. Since the target ratings of the GRASP framework are ordinal, the correlation testing is an appropriate method for showing the interrater reliability. The Spearman's rank correlation coefficient is the best nonparametric correlation estimator. It is widely used in the applied sciences and reported to be a robust measure of correlation [31]. After grading the tools, the two independent researchers were asked to provide their open-ended feedback. Through a short five questions survey, they were asked if the GRASP framework design was logical, if they found it useful, easy to use, their opinion in the criteria used for grading, and if they wish to add, remove, or change any of them.

## 2.4. Analysis and Outcomes

Three major outcomes were planned. Firstly, through the eight closed-ended agreement questions of the survey, the average scores and distributions of experts' opinions on the different criteria used by the GRASP framework to grade and assess the predictive tools should help to improve such criteria. A five-points Likert scale ranging from strongly agree to strongly disagree was used, where the first was assigned the score of five and the last was assigned the score of one, to translate qualitative values into quantitative measures for the sake of the analysis [32, 33]. Secondly, the six open-ended free text questions should provide experts with the opportunity to suggest adding, removing or updating any of the framework criteria. The qualitative analysis should help categorise such suggestions and recommendations into specific information and should also support updating the framework design and detailed content. The qualitative data analysis was conducted using the NVivo Version 12.3 software package [34]. Thirdly, an interrater reliability testing was designed to measure how accurate and consistent the grading of eight predictive tools, conducted by two independent researchers, compared to each other and compared to the grading of the same tools by the authors.

## 3. Results

The literature search generated a list of 1,186 relevant publications. A total of 882 unique emails were identified and extracted from the publications. In six weeks; from the 4th of October to the 15th of November 2018, a total of eighty-one valid responses were received from international experts, with a response rate of 9.2%.

### 3.1. Experts Agreement on GRASP Criteria

The overall average agreement of the eighty-one respondents to the eight closed-ended questions was 4.35; which means the respondents strongly agreed, overall, to the criteria of the GRASP framework. Respondents strongly agreed to six of the eight closed-ended agreement questions, regarding the criteria used by the GRASP framework for evaluating predictive tools. They somewhat agreed to one, and were neutral about another one, of the eight closed-ended agreement questions. Table 1 shows the average agreements of the respondents on each of the eight closed-ended questions and Figure 2 shows the averages and distributions of respondents' agreements on each question. The country distributions of the respondents are shown in Table 6 in the Appendix.

Table 1: Average Agreements of Expert Respondents on Framework Criteria

| SN | Question | Score | Meaning |
|---|---|---|---|
| 1 | Predictive Performance: We should consider the evidence on validating the tool's predictive performance. | 4.87 | Strongly Agree |
| 2 | Predictive Performance Level: The evidence level could be High (internal + multiple external validation), Medium (internal + external validation once), or Low (internal validation only). | 4.44 | Strongly Agree |
| 3 | Usability: We should consider the evidence on the tool's usability. | 4.68 | Strongly Agree |
| 4 | Potential Effect: We should consider the evidence on the tool's potential effect. | 4.61 | Strongly Agree |
| 5 | Usability is Higher: The evidence level on tools' usability should be considered higher than the evidence level on tools' potential effect. | 2.97 | Neither Agree nor Disagree |
| 6 | Impact: We should consider the evidence on the tool's impact on healthcare effectiveness, efficiency or safety. | 4.78 | Strongly Agree |
| 7 | Impact Level: The evidence level could be High (based on experimental studies), Medium (observational studies), or Low (subjective studies). | 4.16 | Somewhat Agree |
| 8 | Evidence Direction: Based on the conclusions of published studies, the overall evidence direction could be Positive, Negative or Mixed. | 4.26 | Strongly Agree |
| **Overall Average** | | **4.35** | **Strongly Agree** |

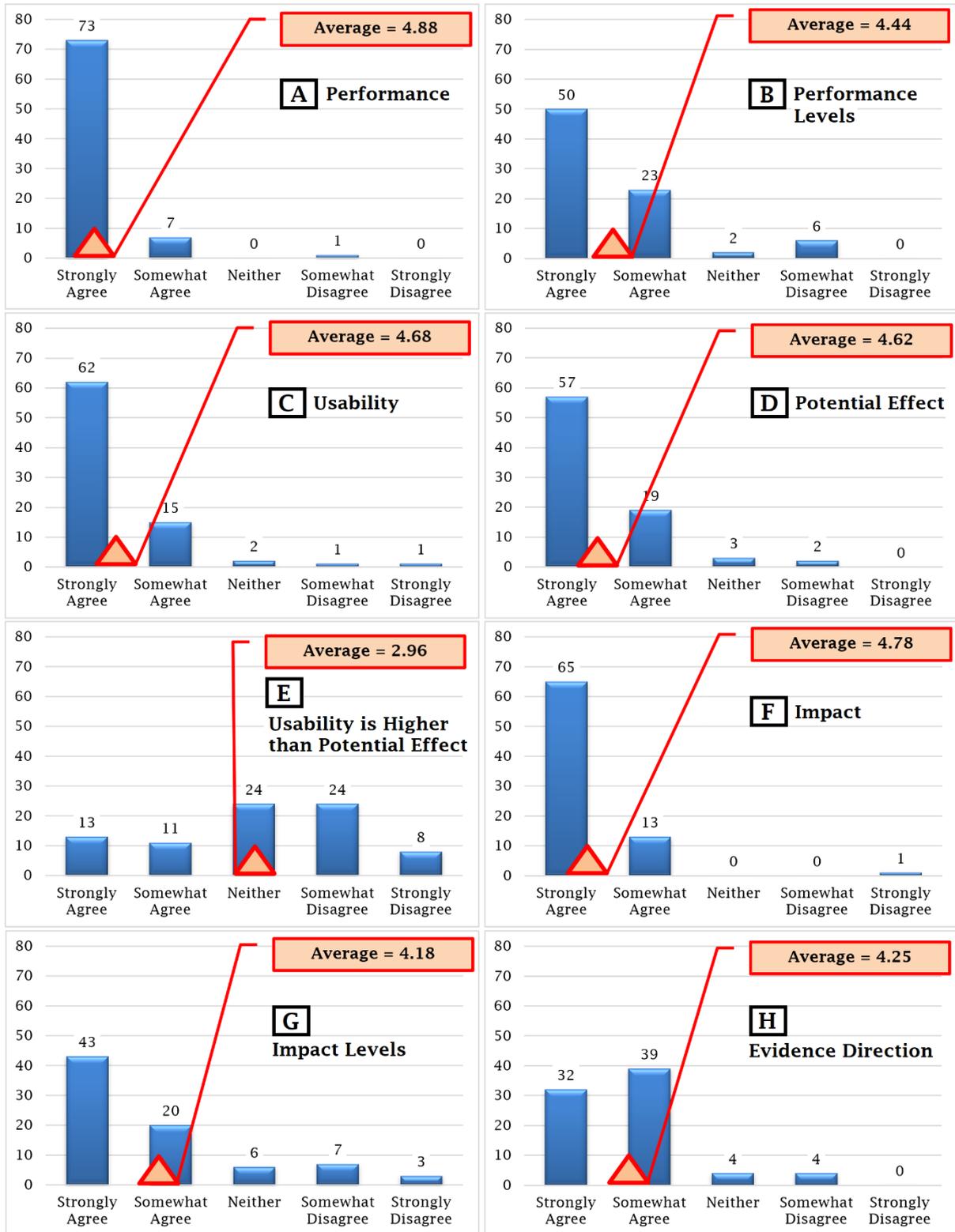

Figure 2: Averages & distributions of respondents' agreements: A - Predictive Performance, B - Performance Levels, C - Usability, D - Potential Effect, E - Usability is Higher than Potential Effect, F - Impact, G - Impact Levels, and H - Evidence Direction

## 3.2. Experts Comments, Suggestions and Recommendations

While the total valid responses were eighty-one; two thirds of the respondents, on average, provided their suggestions or discussed some recommendations for each of the six open-ended free text questions. These questions asked experts for their feedback regarding adding, removing or changing any of the GRASP framework evaluation criteria, their feedback regarding defining and capturing successful tools' predictive performance, when different clinical predictive tasks have different predictive requirements, and their feedback regarding managing conflicting evidence of studies while there is variability in the quality and specifications of published evidence.

### 3.2.1. Predictive Performance and Performance Levels

The respondents discussed that the method, type, and quality of internal and external validation studies should be reported in the GRASP framework detailed report. When external validation studies are conducted multiple times using different patient populations, in different healthcare settings, at different institutions, in different countries, over different times, or by different researchers then the tool is said to have a broad validation range, which means it is more reliable to be used across these different variations of healthcare settings. The respondents said that the tool's predictive performance is considered stable and reliable, when multiple external validation studies produce homogeneous predictive performances, e.g. similar sensitivities and specificities. They also discussed adding the concept of "Strength of Evidence"; which should be mainly based on the quality of the reported study and how much the conditions of the study are close to the original specifications of the predictive tool, in terms of clinical area, population, and target outcomes. It should be part of the components of deciding the direction of evidence (positive, negative, or mixed). It should also be reported in the detailed GRASP framework report, so that users can consider when selecting among two or more tools of the same assigned grade. For example, two predictive tools are assigned grade C1 (each was externally validated multiple times) but one of them shows a strong positive evidence and the other shows a medium or weak positive evidence. It is logic to select the tool with the stronger evidence, if both have similar predictive performances for the same tasks.

### 3.2.2. Usability and Potential Effect

The respondents discussed that the methods and quality of the usability studies and the potential effect studies should be reported in the GRASP framework detailed report. Some of the respondents discussed that the potential effect and usability are not measured during implementation, rather they are measured during the planning for implementation, which is before wide-scale implementation. They also suggested that the details on the potential effect should report the focus on clinical patient outcomes, healthcare outcomes, or provider behaviour. Most of the respondents said that the potential effect is more important than the usability and should have a higher evidence level. A highly usable tool that has no potential effect on healthcare is useless, while a less usable tool that has a promising potential effect is surely better. Some respondents discussed that evaluating both the potential effect and the usability should be considered together as a higher evidence than any of them alone.

### *3.2.3. Post-Implementation Impact and Impact Levels*

The respondents discussed that the method and quality of the post-implementation impact study should be reported in the GRASP framework detailed report. Again, respondents discussed adding the concept of "Strength of Evidence". Within each evidence level of the post-implementation impact we could have several sub-levels, or at least a classification of the quality of studies. for example, not all observational studies are equal in quality; a case series would be very different to a case control or large-scale prospective cohort study. Within the experimental studies we could also have different sub-levels of evidence, quasi-experimental vs. randomised controlled trial for example. These sub-levels should be included in the GRASP framework detailed report, when reporting the individual studies, this will provide the reader with more details on the strength and quality of the evidence on the tools.

### *3.2.4. Direction of Evidence*

Respondents discussed that the direction of evidence should consider the quality and strength of evidence. Most respondents here used the terms; "quality of evidence" and "strength of evidence", synonymously. Respondents discussed that quality of evidence or the strength of evidence should consider many elements of the published study, such as the methods used, the appropriate population, appropriate settings, the clinical practice, the sample size, the type of data collection; retrospective vs prospective, the outcomes, the institute of study and any other quality measures. The direction of evidence depends largely on the quality of the evidence, in case there are conflicting conclusions from multiple studies.

### 3.2.5. Defining and Capturing Predictive Performance

Respondents discussed that the predictive performance evaluation depends basically on the intended prediction task, so this is different from one tool to another, based on the task that each tool does. The clinical condition under prediction and the cost-effectiveness of treatment would highly influence the predictive performance evaluation. Predictive performance evaluation depends also on the actions recommended based on the tool. For example, screening tools should perform with high sensitivity, high negative predictive value, and low likelihood ratio, since there is a following level of checking by clinicians or other tests, while diagnostic tools should always perform with high specificity, high positive predictive value, and high likelihood ratio, since the decisions are based here directly on the outcomes of the tool, and some of these decisions might be risky to the patient or expensive to the healthcare organisation. Respondents discussed that for diagnostic tools, predictive performance is more likely to be expressed through sensitivity and specificity, while for prognostic tools, it is better to express predictive performance through probability/risk estimation. Predictive tools must always be adjusted to the settings, populations, and the intended tasks before their adoption and implementation in the clinical practice.

### 3.2.6. Managing Conflicting Evidence

Respondents discussed that deciding on the conflicting evidence should consider the quality of each study or the strength of evidence, to decide on the overall direction of evidence. Measures include the proper methods used in the study, if the population is appropriate, if the settings are appropriate, if the study is conducted at the clinical practice, if the sample size is large, if the data collection was prospective not retrospective, if the outcomes are clearly reported, if the institute of the study is credible, if the study involved multiple sites or hospitals, and any other quality measures related to the methods or the data. We should rely primarily on conclusions from high-quality low risk of bias studies, as recommended in other fields, e.g. systematic reviews. A well designed and conducted study should have more credibility than a poorly designed and conducted study. If different results are obtained for sub-populations, this should be further investigated and explained. The predictive tool may only perform well in certain sub-populations, based on the intended tasks. If we have evidence from settings outside the target population of the tool, then these shouldn't have much weight, or less weight, on the evidence to support the tool, such as non-equivalent studies; which are conducted to validate a tool for a different population, predictive task, or clinical settings. Much of the important information is in the details of the evidence variability. So, it is important to report this in the framework detailed report, to provide as much details as possible for each reported study to help end users make more accurate decisions based on their own settings, intended tasks, target populations, practice priorities, and improvement objectives.

### *3.3. Updating the GRASP Framework*

Based on the respondents' feedback, on both the closed-ended evaluation criteria agreement questions and the open-ended suggestions and recommendations questions, the GRASP framework concept was updated, as shown in Figure 3. Regarding Phase C; the pre-implementation phase including the evidence on predictive performance evaluation, the three levels of internal validation, external validation once, and external validation multiple times, were additionally assigned "Low Evidence", "Medium Evidence", and "High Evidence" labels respectively. Phase B; During Implementation, has been renamed to "Planning for Implementation". The Potential Effect is now made of higher evidence level than Usability and the evidence of both potential effect and usability together is higher than any one of them alone. Now we have three levels of evidence; B1 = both potential effect and usability are reported, B2 = Potential effect evaluation is reported, and B3 = Usability testing is reported. Figure 4 in the Appendix shows a clean copy of the updated GRASP framework concept.

The GRASP framework detailed report was also updated, as shown in Table 4 in the Appendix. More details were added to the predictive tools information section, such as the internal validation method, dedicated support of research networks, programs, or professional groups, the total citations of the tool, number of studies discussing the tool, the number of authors, sample size used to develop the tool, the name of the journal which published the tool and its impact factor. Table 5 in the Appendix shows the Evidence Summary. This summary table provides users with more information in a structured format on each study discussing the tools, whether these were studies of predictive performance, usability, potential effect or post-implementation impact. Information includes study name, country, year of development, and phase of evaluation. The evidence summary provides more quality related information, such as the study methods, the population and sample size, settings, practice, data collection method, and study outcomes. Furthermore, the evidence summary provides information on the strength of evidence and a label, to highlight the most prominent or important predictive functions, potential effects or post-implementation impacts of the tools.

We developed a new protocol to decide on the strength of evidence. The strength of evidence protocol considers two main criteria of the published studies. Firstly, it considers the degree of matching between the evaluation study conditions and the original tool specifications, in terms of the predictive task, target outcomes, intended use and users, clinical specialty, healthcare settings, target population, and age group. Secondly, it considers the quality of the study, in terms of the sample size, data collection, study methods, and credibility of institute and authors. Based on these two criteria, the strength of evidence is classified into 1) strong evidence: matching evidence of high quality, 2) medium evidence: matching evidence of low quality or non-matching evidence of high quality, and 3) weak evidence: non-matching evidence of low quality. Figure 7 in the Appendix shows the strength of evidence protocol.

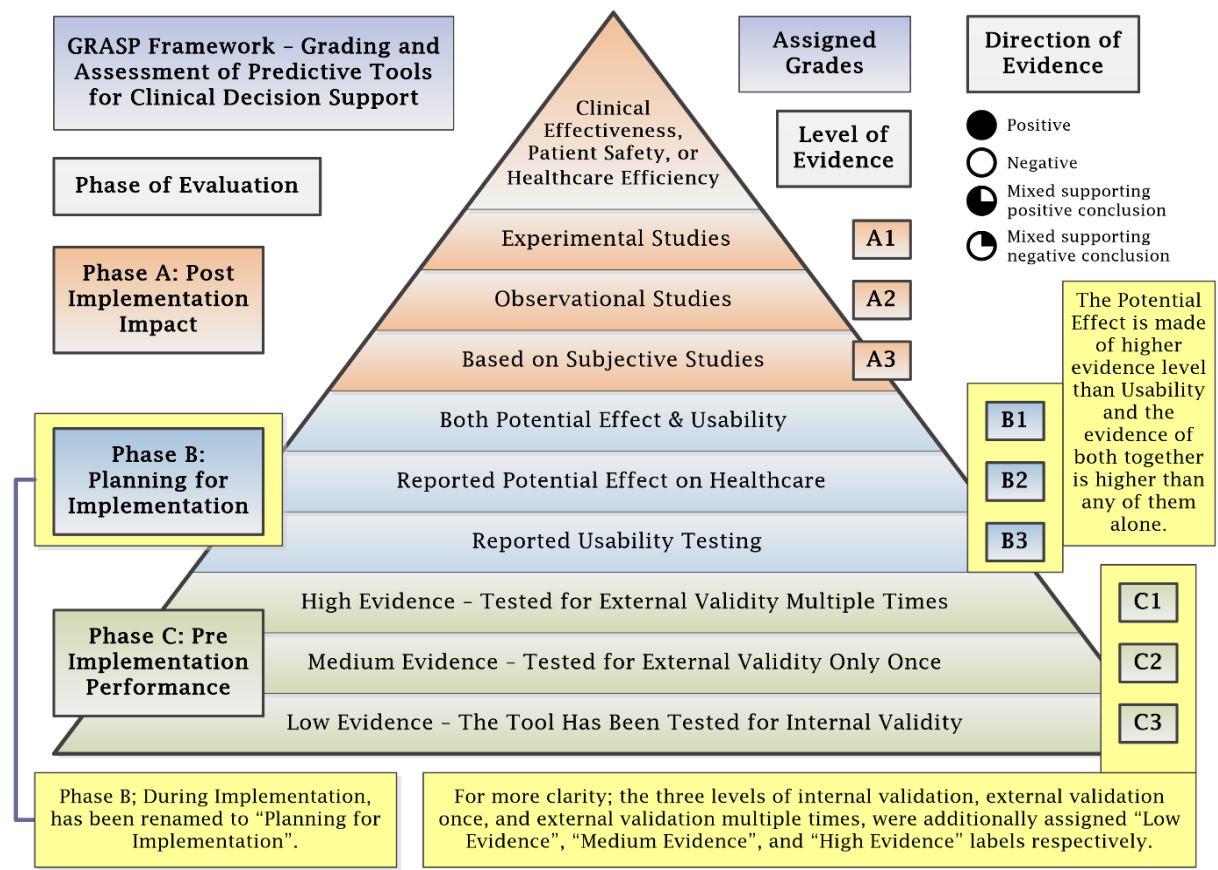

Figure 3: The Updated GRASP Framework Concept

### 3.4. The GRASP Framework Reliability

The two independent researchers assigned grades to the eight predictive tools and produced a detailed report on each one of them. The summary of the two independent researchers assigned grades, compared to the authors, are shown in Table 2. A more detailed information on the justification of the assigned grades is shown in the Appendix in Table 7. The Spearman's rank correlation coefficient was 0.994 (p<0.001) comparing the first researcher to the authors, 0.994 (p<0.001) comparing the second researcher to the authors, and 0.988 (p<0.001) comparing the two researchers to each other. This shows a statistically significant and strong correlation, indicating a strong interrater reliability of the GRASP framework. Accordingly, the GRASP framework produced reliable and consistent grades when it was used by independent users. Both independent researchers found GRASP framework design logical, easy to understand, and well organized. They both found GRASP useful, considering the variability of tools' quality and levels of evidence. They both found it easy to use. They both thought the criteria used for grading were logical, clear, and well structured. They did not wish to add, remove, or change any of the criteria. However, they asked for adding some definitions and clarifications to the criteria, which was included in the update.

Table 2: Grades Assigned by the Two Independent Researchers and the Authors

| Tools | Grading by Researcher 1 | Grading by Researcher 2 | Grading by Authors |
|---|---|---|---|
| Centor Score [35] | **B2** | **B3** | **B3** |
| CHALICE Rule [36] | **B2** | **B2** | **B2** |
| Dietrich Rule [37] | **C0** | **C0** | **C0** |
| LACE Index [38] | **C1** | **C1** | **C1** |
| Manuck Scoring System [39] | **C2** | **C2** | **C2** |
| Ottawa Knee Rule [40] | **A1** | **A2** | **A1** |
| PECARN Rule [41] | **A2** | **A2** | **A2** |
| Taylor Mortality Model [42] | **C3** | **C3** | **C3** |

## 4. Discussion and Conclusion

### *4.1. Brief Summary*

It is a challenging task for most clinicians to critically evaluate a growing number of predictive tools, proposed in the literature, in order to select effective tools for implementation at their clinical practice or for recommendation in clinical practice guidelines, to be used by other clinicians. Although most of these predictive tools have been assessed for predictive performance, only a few have been implemented and evaluated for comparative effectiveness or post-implementation impact. Clinicians need an evidence-based approach to provide them with standardised objective information on predictive tools to support their search for and selection of effective tools for their clinical tasks. Based on the critical appraisal of the published evidence on predictive tools, the GRASP framework uses three dimensions to grade predictive tools: 1) Phase of Evaluation, 2) Level of Evidence and 3) Direction of Evidence. The final grade assigned to a tool is based on the highest phase of evaluation, supported by the highest level of positive evidence, or mixed evidence that supports a positive conclusion. In this paper, we present the validation of the GRASP framework through the feedback of a wide international group of experts, where the GRASP framework concept, evaluation criteria, and the detailed report have been updated based on their feedback. The reliability testing showed that the GRASP framework can be used reliably and consistently by independent users to grade predictive tools.

*4.2. Predictive Performance*

The internal validation of the predictive performance of a tool is essential to make sure the tool is doing the prediction task as designed [43, 44]. The predictive performance is evaluated using measures of discrimination and calibration [45]. While discrimination refers to the ability of the tool to distinguish between patients with and without the outcome under consideration, calibration refers to the accuracy of the prediction, and show how much the predicted and the observed outcomes agree [46]. Discrimination is usually measured through sensitivity, specificity, and the area under the curve (AUC) [47]. On the other hand, calibration could be summarised using the Hosmer-Lemeshow test or the Brier score [48]. The external validation of predictive tools is essential to reflect the reliability and generalisability of the tools [49]. Predictive tools are more reliable and trustworthy, not only when their predictive performance is better, but more importantly when they undergo high quality, multiple, and wide range external validation [50]. The high quality is usually reflected in the type and size of data samples used in the validation, while repeating the external validations on different patient populations, at different institutions, in different healthcare settings, and by different researchers shows higher reliability of the predictive tools [44].

*4.3. Usability and Potential Effect*

In addition to the predictive performance, clinicians are usually interested to learn about the potential effects of the tools on improving patient outcomes, saving time, costs and resources, or supporting patient safety [2, 51]. They need to know more about the expected impact of using the tool on different healthcare aspects, processes or outcomes, assuming the tool has been successfully implemented in the clinical practice [52, 53]. If a CDS tool has less potential to improve healthcare processes or clinical outcomes it will not be easily adopted or successfully implemented in the clinical practice [54]. Some clinicians might also be interested to learn about the usability of predictive tools; whether these tools can be used by the specified users to achieve specified and quantifiable objectives in the specified context of use [55, 56]. CDS tools with poor usability will eventually fail, even if they provide the best performance or potential effect on healthcare [7, 57]. Usability includes several measurable criteria, based on the perspectives of the stakeholders, such the mental effort needed, the user attitude, interaction, easiness of use, and acceptability of systems [58, 59]. Usability can also be evaluated through measuring the effectiveness of task management with accuracy and completeness, the efficiency of utilising resources, and the users' satisfaction, comfort with, and positive attitudes towards, the use of the tools [60, 61], in addition to learnability, memorability and freedom of errors [62, 63].

### *4.4. Post-Implementation Impact*

Clinicians are interested to learn about the post-implementation impact of CDS tools, on different healthcare aspects, processes, and outcomes, before they consider their implementation in the clinical practice [64-66]. The most interesting part of the impact studies for clinicians is the effect size of the CDS tools and their direct impact on physicians' performance and patients' outcomes [67, 68]. Clinicians consider that high quality experimental studies, such as randomised controlled trials, are the highest level of evidence, followed by observational well-designed cohort or case-control studies and lastly subjective studies, opinions of respected authorities, and reports of expert committees or panels [69-71]. For many years, experimental methods have been viewed as the gold standard for evaluation, while observational methods were considered to have little or no value. However, this ignores the limitations of randomised controlled trials, which may prove unnecessary, inappropriate, inadequate, or sometimes impossible. Furthermore, high-quality observational studies have an important role in comparative effectiveness research because they can address issues that are otherwise difficult or impossible to study. Therefore, we need to understand the complementary roles of the two approaches and appreciate the scientific rigour in evaluation, regardless of the method used [72, 73].

### *4.5. Direction of Evidence and Conflicting Conclusions*

It is not uncommon to encounter conflicting conclusions when a predictive tool is validated or implemented and evaluated in different patient subpopulations or for different prediction tasks or outcomes [74, 75]. The cut-off value that determines what a good predictive performance is, for example, depends not only on the clinical condition under consideration but largely on the requirements, conditions, and consequences of the decisions made accordingly [76]. One of the main challenges here is dealing with the huge variability in the quality, types, and conditions of studies published in the literature. This variability makes it impossible to synthesise different measures of predictive performance, usability, potential effect or post-implementation impact into simple quantitative values, like in meta-analysis or systematic reviews [77, 78].

*4.6. The GRASP Framework Overall*

The grades assigned to predictive tools, using the GRASP framework, provide relevant evidence-based information to guide the selection of predictive tools for clinical decision support. However, the framework is not meant to be precisely prescriptive. An A1 tool is not always and absolutely better than an A2 tool. A clinician may prefer an A2 tool showing improved patient safety in two observational studies rather than an A1 tool showing reduced healthcare costs in three experimental studies. It all depends on the objectives and priorities the users are trying to achieve, through implementing and using predictive tools in their clinical practice. More than one predictive tool could be endorsed, in clinical practice guidelines, each supported by its requirements and conditions of use and recommended for its most prominent outcome of predictive performance, potential effect, or post-implementation impact on healthcare and clinical outcomes. The GRASP framework remains a high-level approach to provide clinicians with an evidence-based and comprehensive, yet simple and feasible, method to evaluate and select predictive tools. However, when clinicians need further information, the framework detailed report provides them with the required details to support their decision making.

*4.7. Challenges, Limitations, and Future Work*

It might be easy to analyse the feedback of experts using closed-ended questions. However, analysing the feedback of experts using open-ended questions is rather difficult [79]. Qualitative content and thematic analysis of free text feedback is challenging, since the extraction of significance becomes more difficult with diverse opinions, different experiences, and variable perspectives [80]. It is advised by many healthcare researchers to use Delphi techniques to reach to consensus among experts, through successive rounds of feedback, when developing clinical guidelines or selecting evaluation criteria and indicators [81-83]. However, using the Delphi techniques was not feasible in our study.

Even though we contacted a large number of 882 experts, in the area of developing, implementing and evaluating predictive tools and CDS systems, we got a very low response rate of 9.2%, and received only 81 valid responses. This low response rate could have been improved if participants were motivated by some incentives, more than just acknowledging their participation in the study, or if more support was provided through the organisations these participants belong to, which needs much more resources to synchronise these efforts. For the sake of keeping the survey feasible for most busy experts, the number of the closed ended as well as the open-ended questions were kept limited and the required time to complete the whole survey was kept in the range of 20 minutes. However, some of the participants could have been willing to provide more detailed feedback, through interviews for example, which was out of the scope of this study and was not initially possible to conduct with all the invited experts, otherwise we would have received a much lower response rate.

To evaluate the impact of the GRASP framework on clinicians' decisions and examine the application of the framework to grade predictive tools, the authors are currently working on two more studies. The first study should validate and evaluate the impact of using the framework on improving the decisions made by end user clinicians and healthcare professionals, regarding selecting predictive tools for the clinical tasks. Through an online survey of a wide international group of clinicians and healthcare professionals, the study should compare the performance and outcomes of making decisions with and without using the framework. The second study aims to apply the framework to a large consistent group of predictive tools, used for the same clinical prediction task. This study should show how the framework provides clinicians with an evidence-based method to compare, evaluate, and select predictive tools, through grading and reporting tools based on the critical appraisal of their published evidence.

To enable end user clinicians and clinical practice guideline developers to access detailed information, reported evidence and assigned grades of predictive tools, it is essential to discuss implementing the GRASP framework into an online platform. However, maintaining such grading system up to date is a challenging task, as this requires the continuous updating of the predictive tools grading and assessments, when new evidence becomes published and available. It is important to discuss using automated or semi-automated methods for searching and processing new information to keep the GRASP framework updated. Finally, we recommend that the GRASP framework be utilised by working groups of professional organisations to grade predictive tools, in order to provide consistent results and increase reliability and credibility for end users. These professional organisations should also support disseminating such evidence-based information on predictive tools, in a similar way of announcing and disseminating new updates of clinical practice guidelines.

## 5. Declarations

**Acknowledgments**

We would like to thank all the professors, doctors, and researchers who participated in the validation of the GRASP framework including; Abdullah Pandor, Adam Dunn, Alberto Zamora Cervantes, Alex C Spyropoulos, Allyson R Cochran, Alyson Mahar, Anders Granholm, Andrew D MacCormick, Anupam Kharbanda, Ashraf El-Metwally, Beth Devine, Brian Shirts, Carme Carrion, Carrie Ritchie, Cesar Garriga, Christoph U Lehmann, Claudia Gasparini, Claudia Pagliari, Craig Anderson, Douglas P. Gross, Dustin Ballard, Erik Roelofs, Ewout W. Steyerberg, Fabian Jaimes, Felix Zubia-Olaskoaga, Fernando Ferrero, Gary Collins, Gary Maartens, Grégoire Le Gal, Ilkka Kunnamo, Janneke Stalenhoef, Jitendra Jonnagaddala, Julian Brunner, Kent P. Hymel, Kristen Miller, Laura Cowley, Liliana Laranjo da Silva, Luke Daines, Manish Kharche, Maria Lourdes Posadas-Martinez, Mark Ebell, Maryati Mohd. Yusof, Matthias Döring, Matthijs Becker, Maxwell Dalaba, Michael T Weaver, Michelle Ng Gong, Mohamed Hassan Ahmed Fouad, Mowafa Househ, NadÃ¨ge Lemeunier, Natalie Edelman, Nathan Dean, Nick van Es, Omar S. Al-Kadi, Oscar Perez Concha, Patrick Vanderstuyft, Peter Dayan, Peter Kent, Pieter Cornu, Rabia Bashir, Reza Khajouei, Robert C Amland, Robert E. Freundlich, Robert Greenes, Rose Galvin, Samina Abidi, Seong Ho Park, Sheila Payne, Sherif Shabana, Simon Adams, Surbhi Leekha, Syed Mustafa Ali, Tero Shemeikka, Thomas Debray, Vassilis Koutkias.

**Funding**

This work was supported by the Commonwealth Government Funded Research Training Program, Australia.

**Authors' contributions**

MK mainly contributed to the conception, detailed design, and conduction of the study. BG and FM supervised the study from the scientific perspective. BG was responsible for the overall supervision of the work done, while FM was responsible for providing advice on the enhancement of the methodology used and data analysis conducted. All the authors have been involved in drafting the manuscript and revising it. Finally, all the authors gave approval of the manuscript to be published and agreed to be accountable for all aspects of the work.

**Ethics approval and consent to participate**

This study has been approved by the Human Research Ethics Committee, Faculty of Medicine and Health Sciences, Macquarie University, Sydney, Australia, on the 4th of October 2018. Reference No: 5201834324569. Project ID: 3432.

**Consent to publication**

Not applicable.

**Competing interests**

The authors declare that they have no competing interests.

**Author details**

[1] Australian Institute of Health Innovation, Faculty of Medicine and Health Sciences, Macquarie University, 75 Talavera Road, North Ryde, Sydney, NSW 2113, Australia.

[2] Centre for Big Data Research in Health, University of New South Wales, Kensington, Sydney, NSW 2052, Australia.

# 7. The Appendix

## 7.1. The GRASP Framework Detailed Report

Table 3: The GRASP Framework Detailed Report

| Name | Name of predictive tool (report tool's creators and year in the absence of a given name) | | | | | | | |
|---|---|---|---|---|---|---|---|---|
| Authors/Year | Name of developer, country and year of publication | | | | | | | |
| Intended use | Predictive task/specific aim/intended use of the predictive tool | | | | | | | |
| Intended user | Type of practitioner intended to use the tool (e.g. physician or nurse) | | | | | | | |
| Category | Diagnostic/Therapeutic/Prognostic/Preventive | | | | | | | |
| Clinical area | Clinical specialty | | | | | | | |
| Target Population | Target patient population and health care settings in which the tool is applied | | | | | | | |
| Target Outcome | Event to be predicted (including prediction lead time if needed) | | | | | | | |
| Action | Recommended action based on tool's output | | | | | | | |
| Input source | • Clinical (including Diagnostic, Genetic, Vital signs, Pathology)<br>• Non-Clinical (including Healthcare Utilisation) | | | | | | | |
| Input type | • Objective (Measured input; from electronic systems or clinical examination)<br>• Subjective (Patient reported; history, checklist …etc.) | | | | | | | |
| Local context | Is the tool developed using location-specific data? (e.g. life expectancy tables) | | | | | | | |
| Methodology | Type of algorithm (e.g. parametric/non-parametric) | | | | | | | |
| Endorsement | Organisations endorsing the tool and/or guidelines recommending its utilisation | | | | | | | |
| Automation Flag | Automation status (manual/automated) | | | | | | | |
| **Phase of Evaluation** | **Level of Evidence** | **Grade** | **Evaluation Studies** | | | | | |
| Phase C: Before implementation Is it possible? | Insufficient internal validation | C0 | Tested for internally validity but was either insufficiently internally validated or validation was insufficiently reported. | | | | | |
| | Internal validation | C3 | Tested for internally validity (reported calibration & discrimination; sensitivity, specificity, positive and negative predictive values & other performance measures). | | | | | |
| | External validation | C2 | Tested for external validity, using one external dataset. | | | | | |
| | External validation multiple times | C1 | Tested multiple times for external validity, using more than one external dataset. | | | | | |
| Phase B: During implementation Is it practicable? | Potential effect | B2 | Reported estimated potential effect on clinical effectiveness, patient safety or healthcare efficiency. | | | | | |
| | Usability | B1 | Reported usability testing (effectiveness, efficiency, satisfaction, learnability, memorability, and minimizing errors). | | | | | |
| Phase A: After implementation: Is it desirable? | Evaluation of post implementation impact on Clinical Effectiveness, Patient Safety or Healthcare Efficiency | A3 | Based on subjective studies; e.g. the opinion of a respected authority, clinical experience, a descriptive study, or a report of an expert committee or panel. | | | | | |
| | | A2 | Based on observational studies; e.g. a well-designed cohort or case-control study. | | | | | |
| | | A1 | Based on experimental studies; properly designed, widely applied randomised/nonrandomised controlled trial. | | | | | |
| Final Grade | Grade ABC,123 | | A1 | A2 | A3 | B1 | B2 | C1 | C2 | C3 |
| Direction of Evidence | ● Positive Evidence    ◐ Mixed Evidence Supporting Positive Conclusion<br>○ Negative Evidence    ◓ Mixed Evidence Supporting Negative Conclusion | | | | | | | |
| Justification | Explains how the final grade is assigned based on evidence; which conclusions were taken into consideration, as positive evidence, and which were considered negative. | | | | | | | |
| References | Details of studies supporting justification: phase of evaluation, level of evidence, direction of evidence, study type, study settings, methodology, results, findings and conclusions (highlighted according to colour code). | | | | | | These two sections are included in the full GRASP report on each tool. | |
| Label/Colour Code | • Positive Findings<br>• Negative Findings | | • Important Findings<br>• Less Relevant Findings | | | | | |

## 7.2. The Updated GRASP Framework Concept

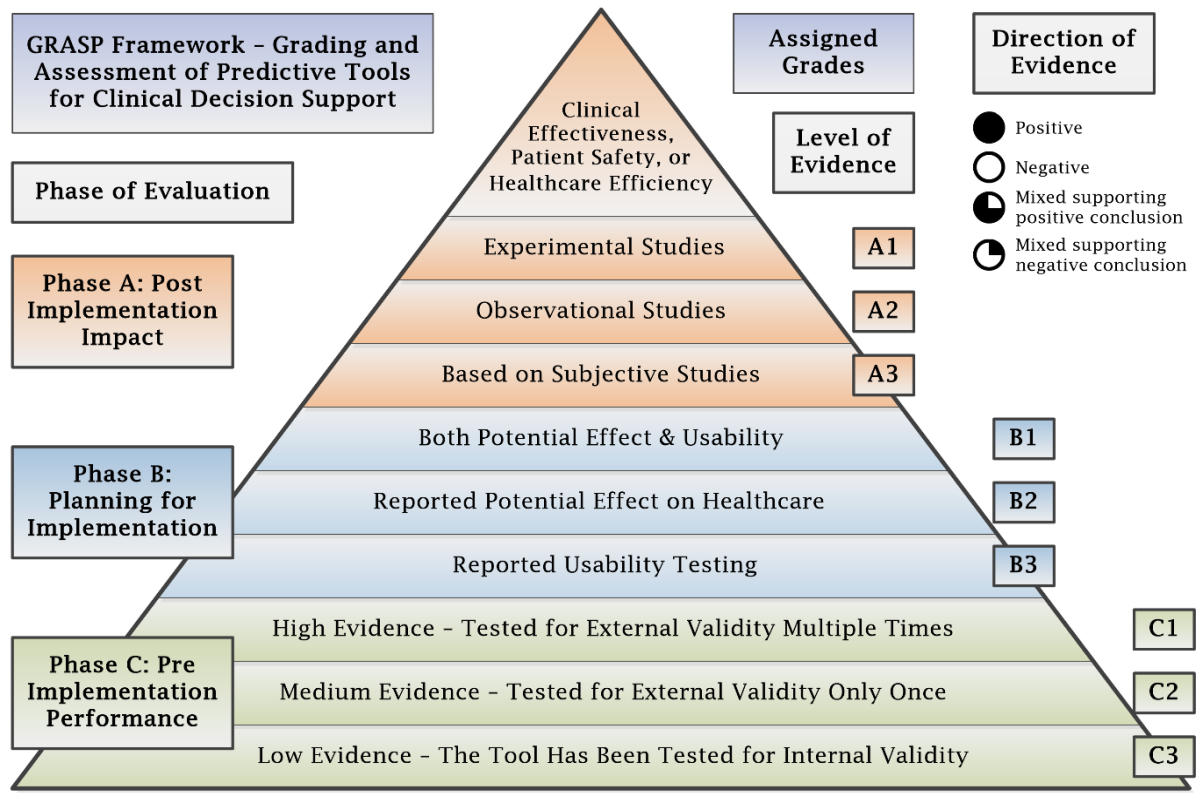

Figure 4: The Updated GRASP Framework Concept

## 7.3. The Updated GRASP Framework Detailed Report

Table 4: The Updated GRASP Framework Detailed Report

| Name | Name of predictive tool (report tool's creators and year in the absence of a given name) |
|---|---|
| Author | Name of developer (first author or researcher) |
| Country | Country of development |
| Year | Year of development |
| Category | Diagnostic/Therapeutic/Prognostic/Preventive |
| Intended use | Predictive task/specific aim/intended use of the predictive tool |
| Intended user | Type of practitioner intended to use the tool (e.g. physician or nurse) |
| Clinical area | Clinical specialty |
| Target Population | Target patient population and health care settings in which the tool is applied |
| Target Outcome | Event to be predicted (including prediction lead time if needed) |
| Action | Recommended action based on tool's output |
| Input source | • Clinical (including Diagnostic, Genetic, Vital signs, Pathology)<br>• Non-Clinical (including Healthcare Utilisation) |
| Input type | • Objective (Measured input; from electronic systems or clinical examination)<br>• Subjective (Patient reported; history, checklist ...etc.) |
| Local context | Is the tool developed using location-specific data? (e.g. life expectancy tables) |
| Methodology | Type of algorithm used for developing the tool (e.g. parametric/non-parametric) |
| Internal Validation | Method of internal validation |
| Dedicated Support | Name of the supporting/funding research networks, programs, or professional groups |
| Endorsement | Organisations endorsing the tool and/or clinical guidelines recommending its utilisation |
| Automation Flag | Automation status (manual/automated) |
| Tool Citations | Total citations of the tool |
| Studies | Number of studies reporting the tool |
| Authors No | Number of authors |
| Sample Size | Size of patient/record sample used in the development of the tool |
| Journal Name | Name of the journal that published the tool's primary development study |
| Journal Rank | Impact factor of the journal |
| Citation Index | Calculated as: Average Annual Citations = number of citations/age of primary publication |
| Publication Index | Calculated as: Average Annual Studies = number of studies/age of primary publication |
| Literature Index | Calculated as: Citations and Publications = number of citations X number of studies |

| Phase of Evaluation | Level of Evidence | Grade | Evaluation Studies |
|---|---|---|---|
| Phase C: Before implementation Is it possible? | Insufficient internal validation | C0 | Not tested for internal validity, insufficiently internally validated, or internal validation was insufficiently reported. |
| | Internal validation | C3 | Tested for internally validity (reported calibration & discrimination; sensitivity, specificity, positive and negative predictive values & other performance measures). |
| | External validation | C2 | Tested for external validity, using one external dataset. |
| | External validation multiple times | C1 | Tested multiple times for external validity, using more than one external dataset. |
| Phase B: Planning for implementation Is it practicable? | Usability | B3 | Reported usability testing (tool effectiveness, efficiency, satisfaction, learnability, memorability, and minimizing errors). |
| | Potential effect | B2 | Reported estimated potential effect on clinical effectiveness, patient safety or healthcare efficiency. |
| | Potential effect & Usability | B1 | Both potential effect and usability are reported. |
| Phase A: After implementation: Is it desirable? | Evaluation of post implementation impact on Clinical Effectiveness, Patient Safety or Healthcare Efficiency | A3 | Based on subjective studies; e.g. the opinion of a respected authority, clinical experience, a descriptive study, or a report of an expert committee or panel. |
| | | A2 | Based on observational studies; e.g. a well-designed cohort or case-control study. |
| | | A1 | Based on experimental studies; properly designed, widely applied randomised/nonrandomised controlled trial. |
| Final Grade | Grade ABC/123 | A1  A2  A3  B1  B2  B3  C1  C2  C3 | |
| Tool Label | One-word description of the most prominent prediction, potential effect or impact on healthcare processes or outcomes. E.g. "Grade A2 – Efficiency" (the tool improves efficiency by saving money, resources or time, proved through observational post-implementation impact studies). | | |
| Direction of Evidence | ● Positive Evidence   ◐ Mixed Evidence Supporting Positive Conclusion   ○ Negative Evidence   ◑ Mixed Evidence Supporting Negative Conclusion | | |
| Justification | Explains how the final grade is assigned based on evidence; which conclusions were taken into consideration, as positive evidence, and which were considered negative. | | |
| Evidence Summary | Details of studies; using the Evidence Summary, to support the justification, where comparative predictive performance and effectiveness studies are highlighted. | | |
| Findings Codes | Positive Findings / Negative Findings / Important Findings | | |

## 7.4. The Evidence Summary

Table 5: The Evidence Summary

| | |
|---|---|
| **Study** | The published study (According to Reference Style) |
| **Country** | Country of study |
| **Year** | Year of study |
| **Phase** | Before Implementation, planning for implementation, after Implementation |
| **Type** | Development / Internal Validation / External Validation / Usability / Potential Effect / Post-Implementation Impact. |
| **Tools** | Single Tool vs Comparative Study (comparing multiple tools or one tool vs clinical practice). |
| **Intended use** [1] | Predictive task/specific aim/intended use of the predictive tool |
| **Intended user** [1] | Type of practitioner intended to use the tool (e.g. physician or nurse) |
| **Clinical Area** [1] | Clinical specialty |
| **Target Population** [1] | Patients (age group, gender group, clinical specifications, e.g. cardiac population). Providers (age group, gender, clinical specifications, e.g. specialty). |
| **Settings** [1] | Inpatient, outpatient, intensive care … etc. |
| **Methods** [2] | Tool development methods: recursive partitioning, multivariate logistic regression … etc. Internal validation methods: out-of-sample, bootstrapping, cross validation, split sample … etc. External validation methods: national, international … etc. Usability: acceptance, satisfaction, adoption … etc. Potential Effect: feasibility, cost-effectiveness, economic analysis … etc. Impact: experimental (randomised, non-randomised, controlled, quasi-experimental), observational (cohort studies, case-control, cross-sectional), subjective (expert opinion, reports) … etc. |
| **Practice** [2] | Clinical vs non-clinical practice. |
| **Sample Size** [2] | Number of patients/records/users recruited in the study |
| **Data Collection** [2] | Prospective/retrospective data |
| **Outcomes** [2] | Reported outcome measures: Development/Validation: reported calibration/discrimination; sensitivity, specificity, positive & negative predictive values & other performance measures. Usability: acceptance, satisfaction … etc. Potential Effect: feasibility, cost-effectiveness, economic analysis … etc. Impact: effect size, duration of implementation … etc. |
| **Institute** [2] | Name and type of hospital (Multiple hospitals, single hospital, tertiary care … etc). |
| **Support** [2] | Dedicated support of research networks, programs or groups. |
| **Authors** [2] | Number of researchers. |
| **Journal** [2] | Journal name and impact factor. |
| **Direction of Evidence** | Positive / Equivocal / Negative (Based on study findings and conclusions). |
| **Matching of Evidence** | Considering fields [1] (Matching/Non-Matching to the tool's original specifications) |
| **Quality of Evidence** | Considering fields [2] (High Quality/Low Quality of the study) |
| **Strength of Evidence** | Based on Evidence Matching and Quality: Strong Evidence / Medium Evidence / Weak Evidence |
| **Label** | Effectiveness / Efficiency / Safety / Workflow / Processes (one or more). |
| **Notes** | Special important study information. |

## 7.5. Experts' Country Distributions

Table 6: Country Distributions of Expert Respondents

| Country | Responses | Percent | Cumulative |
|---|---|---|---|
| United States | 20 | 24.7% | 24.7% |
| United Kingdom | 12 | 14.8% | 39.5% |
| Canada | 8 | 9.9% | 49.4% |
| Netherlands | 5 | 6.2% | 55.6% |
| Spain | 5 | 6.2% | 61.7% |
| Australia & New Zealand | 5 | 4.9% | 67.9% |
| Argentina | 2 | 2.5% | 70.4% |
| Belgium | 2 | 2.5% | 72.8% |
| Germany | 2 | 2.5% | 75.3% |
| Europe | 8 | 9.9% | 85.2% |
| Asia | 6 | 7.4% | 92.6% |
| Africa | 2 | 2.5% | 95.1% |
| South America | 2 | 2.5% | 97.5% |
| Middle East | 2 | 2.5% | 100.0% |
| **Total** | **81** | **100%** | |

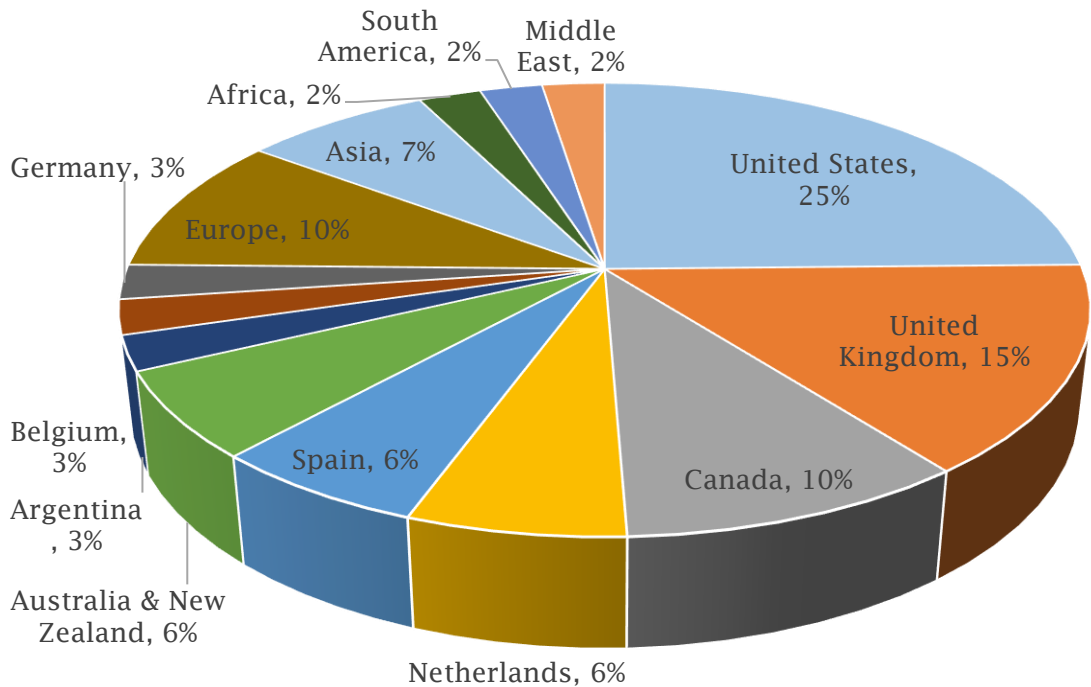

Figure 5: Country Distributions of Expert Respondents

## 7.6. The Mixed Evidence Protocol

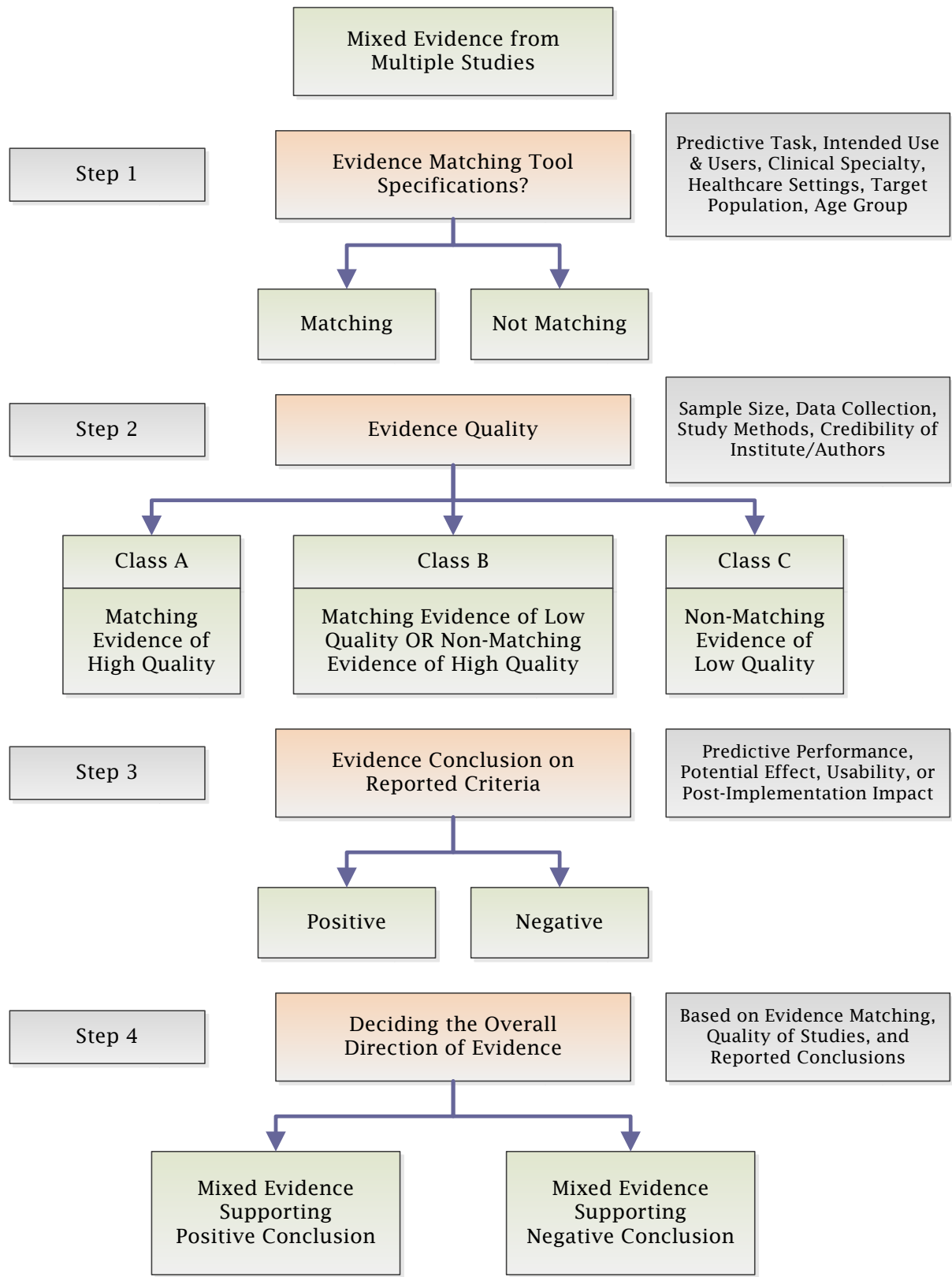

Figure 6: The Mixed Evidence Protocol

**The Mixed Evidence Protocol**

The mixed evidence protocol is based on four steps. Firstly, it considers the degree of matching between the evaluation study conditions and the original tool specifications, in terms of the predictive task, outcome, intended use and users, clinical specialty, healthcare settings, target population, and age group. Secondly, it considers the quality of the study, in terms of sample size, data collection, study methods, and credibility of institute or authors. Based on these two criteria, the studies in the mixed evidence on the tool are classified into 1) Class A: matching evidence of high quality, 2) Class B: matching evidence of low quality or non-matching evidence of high quality, and 3) Class C: non-matching evidence of low quality. Thirdly, it considers the evidence conclusion on the reported evaluation criteria; the predictive performance, potential effect, usability, and post-implementation impact. In the fourth step, studies evaluating predictive tools in closely matching conditions to the tool specifications and providing high quality evidence, Class A, are considered first; taking into account their conclusions on the evaluation criteria in deciding the overall direction of evidence. On the other hand, studies evaluating predictive tools in different conditions to the tool specifications and providing low quality evidence, Class C, are considered last. The conclusion of one study in Class A is considered a stronger evidence than the conflicting conclusions of any number of studies in Class B or C, and the overall direction of the evidence is decided towards the conclusion of the study of Class A. When multiple studies of the same class; for example, Class A, report conflicting conclusions, then we compare the number of studies reporting positive conclusions to those reporting negative conclusions and the overall direction of the evidence is decided towards the conclusion of the larger group. If the two groups are of the same size, then we check if there are more studies in other classes, if not then we examine the reported evaluation criteria and their values in the two groups of studies.

## 7.7. The Strength of Evidence Protocol

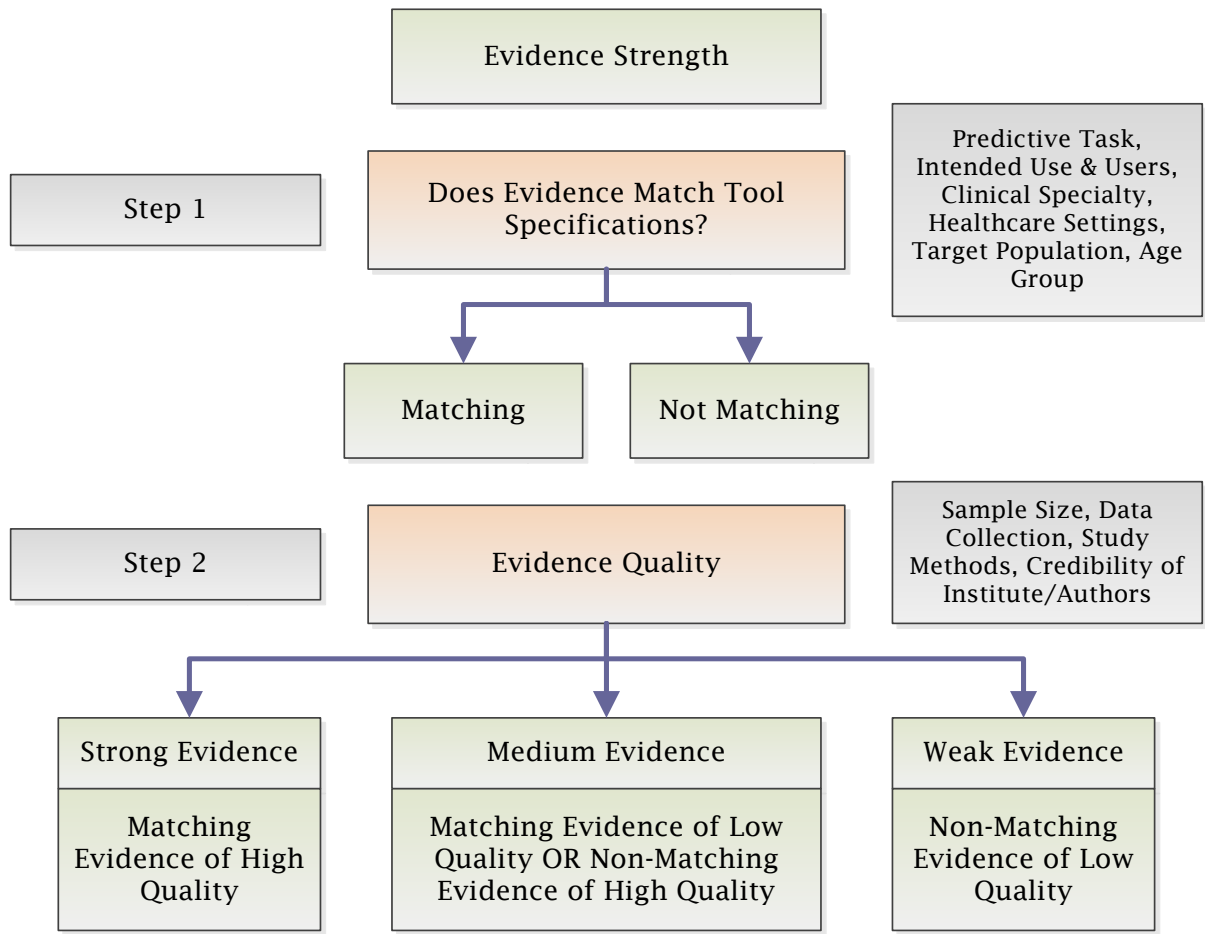

Figure 7: The Strength of Evidence Protocol

## 7.8. Interrater Reliability Detailed Results

Table 7: Grading the Predictive Tools by the Independent Researcher vs the Authors

| Tool | Two Researchers vs Authors | Assigned Grade | Impact After Implementation | | | Planning for Implementation | | | Performance Before Implementation | | |
|---|---|---|---|---|---|---|---|---|---|---|---|
| | | | Experimental Studies | Observational Studies | Subjective Studies | Potential Effect & Usability | Potential Effect | Usability | External Validation Multiple Times | External Validation Only Once | Internal Validation |
| | | | A1 | A2 | A3 | B1 | B2 | B3 | C1 | C2 | C3 |
| Centor Score [35] | R1 | B2 | ○ | | | | ◐ | | ● | | ● |
| | R2 | B3 | ◐ | | | | ○ | ● | ● | | ● |
| | A | B3 | ◐ | | | | ● | ● | ● | | ● |
| CHALICE Rule [36] | R1 | B2 | | | | | ◐ | | ◐ | | ● |
| | R2 | B2 | | | | | ◐ | | ● | | ● |
| | A | B2 | | | | | ◐ | | ● | | ● |
| Dietrich Rule [37] | R1 | C0 | | | | | | | | | ○ |
| | R2 | C0 | | | | | | | | | ○ |
| | A | C0 | | | | | | | | | ○ |
| LACE Index [38] | R1 | C1 | | | | | | | ◐ | | ● |
| | R2 | C1 | | | | | | | ◐ | | ● |
| | A | C1 | | | | | | | ◐ | | ● |
| Manuck Scoring System [39] | R1 | C2 | | | | | | | | ● | ● |
| | R2 | C2 | | | | | | | | ● | ● |
| | A | C2 | | | | | | | | ● | ● |
| Ottawa Knee Rule [40] | R1 | A1 | ● | | | | ● | | ● | | ● |
| | R2 | A2 | | ● | | | | | ● | | ● |
| | A | A1 | ● | | | | | | ● | | ● |
| PECARN Rule [41] | R1 | A2 | | ● | | | | ● | ◐ | | ● |
| | R2 | A2 | | ◐ | | ● | ● | | ● | | ● |
| | A | A2 | | ◐ | | | ● | | ● | | ● |
| Taylor Mortality Model [42] | R1 | C3 | | | | | | | | | ● |
| | R2 | C3 | | | | | | | | | ● |
| | A | C3 | | | | | | | | | ● |
| **Direction of Evidence** | ● Positive Evidence | | | | | ◐ Mixed Evidence Supporting Positive Conclusion | | | | | |
| | ○ Negative Evidence | | | | | ◐ Mixed Evidence Supporting Negative Conclusion | | | | | |

## 7.9. The Survey Screenshots

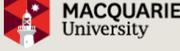

Section 1: The survey introduction

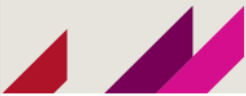

**Published Evidence on Evaluating the Tools Before Implementation**

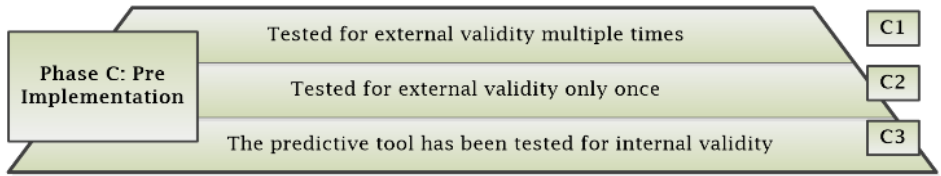

How much do you agree with the following?

|  | Strongly agree | Somewhat agree | Neither agree nor disagree | Somewhat disagree | Strongly disagree |
|---|---|---|---|---|---|
| Q1) We should consider the evidence on validating the tool's predictive performance. | ○ | ○ | ○ | ○ | ○ |
| Q2) The evidence level could be High (internal + multiple external validation), Medium (internal + external validation once), or Low (internal validation only). | ○ | ○ | ○ | ○ | ○ |

Validating the predictive performance includes measures such as sensitivity and specificity of tools. Internal Validation includes testing the performance of the tool on the same data that was used for its development, while External Validation includes testing the performance on new data, different from that used for development.

Q3) Do you suggest adding, removing or changing any of the items or evidence levels?

Section 2: Criteria of evaluating tools' predictive performance before implementation

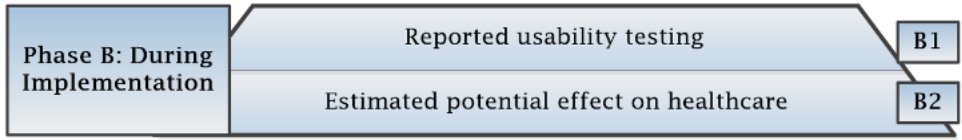

Section 3: Criteria of evaluating tools' usability and estimated potential effect

### Published Evidence on Evaluating the Tools After Implementation

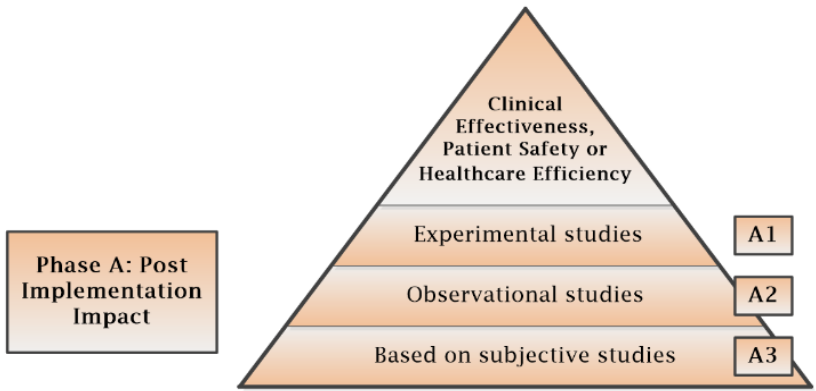

**How much do you agree with the following?**

|  | Strongly agree | Somewhat agree | Neither agree nor disagree | Somewhat disagree | Strongly disagree |
|---|---|---|---|---|---|
| Q8) We should consider the evidence on the tool's impact on healthcare effectiveness, efficiency or safety. | ○ | ○ | ○ | ○ | ○ |
| Q9) The evidence level could be High (based on experimental studies), Medium (observational studies), or Low (subjective studies). | ○ | ○ | ○ | ○ | ○ |

Effectiveness includes improving clinical outcomes, Efficiency includes reducing costs, while Safety includes minimising patient complications. We are assuming that all study types (experimental, observational & subjective) have the same high quality.

Q10) Do you suggest adding, removing or changing any of the items or evidence levels?

Section 4: Criteria of evaluating tools' impact post-implementation

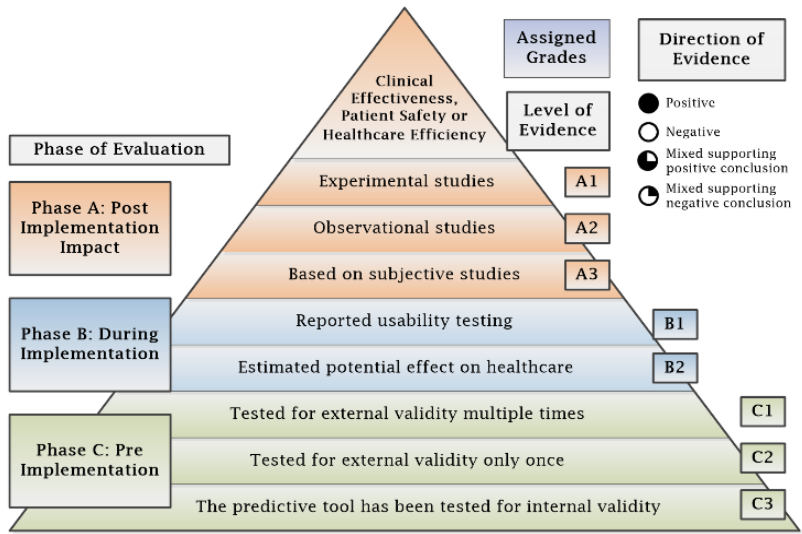

Section 5: Criteria of evaluating direction of published evidence

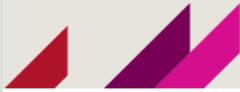

**More Suggestions**

**To develop and improve this framework further:**

**Q13) How would you define and capture successful tools' performance, when different clinical prediction tasks have different predictive performance requirements?**
(e.g. some predictive tasks, such as excluding pulmonary embolism at the emergency department, need high sensitivity while other predictive tasks, such as predicting patients' readmission after discharge, could be achieved with lower sensitivity).

**Q14) How would you manage conflicting evidence of studies while there is variability in the quality and/or sub-populations of the published evidence?**

Section 6: Defining successful predictive performance and managing conflicting evidence.

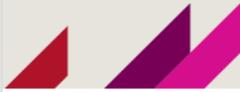

**Feedback and Acknowledgement:**

☐ I would like to receive a feedback on the results of this survey.
☐ I would like to be acknowledged in the publication of this study.

**My name is:**

**My email is:**

All your answers will be kept confidential and will only be used for the purposes of research. Your information is collected only to give you a feedback and will be kept confidential and separate from your answers.

For further information, please contact

Dr. Mohamed Khalifa
Australian Institute of Health Innovation
Macquarie University, 75 Talavera Rd, North Ryde, Sydney, NSW 2113, Australia
M: +61 438 632 060 | E: mohamed.khalifa@mq.edu.au

**Research Chief Investigator:**

A/Prof Blanca Gallego
Australian Institute of Health Innovation
Macquarie University, 75 Talavera Rd, North Ryde, Sydney, NSW 2113, Australia
T: +61 (02) 9850 1608 | E: blanca.gallegoluxan@mq.edu.au

For any complaints about ethical aspects of the research, you can contact:

Ms Vanessa Cooper
Ethics Officer, Australian Institute of Health Innovation
17 Wally's Walk, Level 3, Macquarie University, North Ryde, NSW 2113, Australia
T: +61 (02) 9850 2326 | E: vanessa.cooper@mq.edu.au

Section 7: providing contacts to request feedback and acknowledgment